\documentclass{moriond}

\def\be{\begin{equation}}
\def\ee{\end{equation}}
\def\bea{\begin{eqnarray}}
\def\eea{\end{eqnarray}}

\newcommand{\Photo}{\includegraphics[height=35mm]{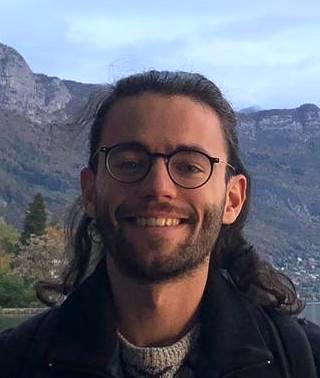}}

\begin{document}
\vspace*{4cm}
\title{Mapping the CMB with QUBIC spectral imaging}

\author{ A. Huchet, T. Laclavère and L. Kardum for the QUBIC collaboration}

\address{Laboratoire Astroparticule et Cosmologie, 10 rue Alice Domon et Léonie Duquet,\\
75205 Paris Cedex 13, France}

\maketitle\abstracts{
QUBIC, the Q \& U Bolometric Interferometer for Cosmology, is a telescope that observes the polarisation of the sky in the millimetre-wavelength range. Its goal is to detect the primordial B-modes of polarisation in the cosmic microwave background by combining the sensitivity of bolometers with the good understanding of interferometry systematics. This dual aspect of QUBIC allows it to perform spectral imaging, that is, obtaining spatial and spectral information of the sky simultaneously. This makes the separation of components with complex spectral energy distributions easier, hence improving the performance of foregrounds removal. We developed three different map making methods (frequency, component and neural network map making) that take advantage of these characteristics. Moreover, QUBIC resumed observing the sky early March and is continuing its commisioning phase with, namely, observations of the Moon.
}

\section{Introduction}

The current standard model of cosmology, $\Lambda$CDM for dark energy Cold Dark Matter, does not describe what happened in the earliest instants of the Universe. One scenario that could explain the intial conditions of $\Lambda$CDM is the ``inflation''. It is a rapidly expanding epoch that stretched quantum variations to cosmological scales. For some inflation models, this would result in the creation of primary polarisation B-modes in the Cosmic Microwave Background (CMB). QUBIC full instrument (FI) will observe a 1.5~\% sky patch in the Southern hemisphere for three years in order to reach the sensitivity necessary to observe those B-modes.

The current version of QUBIC~\cite{QUBIC_I} is a technical demonstrator (TD), with less horns and detectors than the FI. Its location in Argentina on la Puna de Atacama plateau 4869~meters above sea level minimises the effects of atmosphere on the data, but its impact is still important. Moreover, our galaxy (with dust and synchrotron radiation) is another complex foreground that has to be removed through component separation in order to get a pure CMB map. This separation is possible thanks to the different spectral signature of each component.

\section{Spectral imaging with QUBIC}

\begin{figure}
\includegraphics[width=0.55\textwidth]{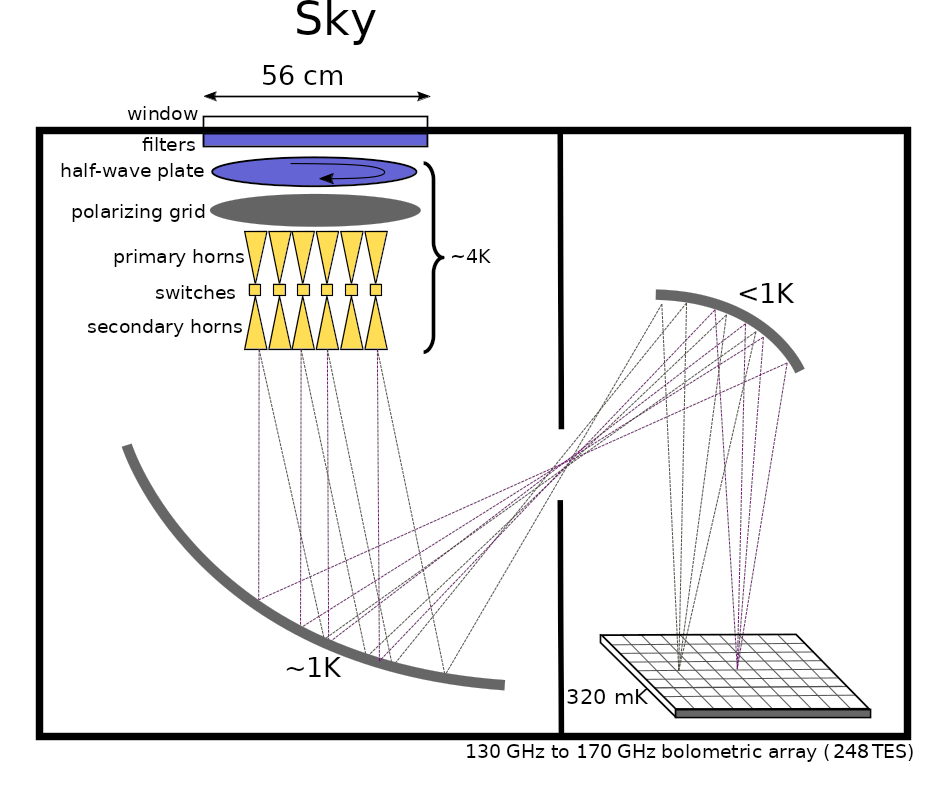}
\includegraphics[width=0.40\textwidth]{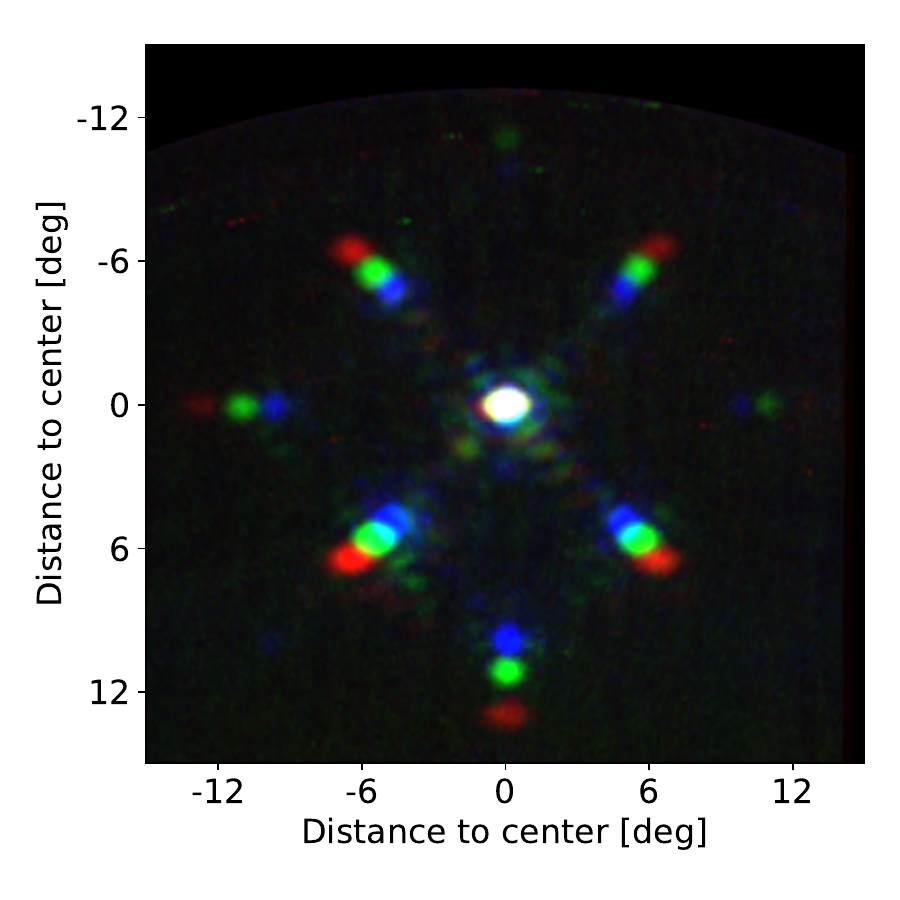}
\caption[]{\textit{Left:} QUBIC Technical Demonstrator schematics, with its 64 back-to-back feedhorn array and 248 transition edge sensors (TES) detector matrix. The instrument has three temperature stages for the optics and detectors: at 4~K, 1~K and 320~mK. \textit{Right:} lab measurement of QUBIC beam for one detector at three frequencies (130, 150, 170 GHz in respectively red, green and blue). The order 0 lobe is easily identified as it is the only one with a position that does not depend on wavelength and is therefore white on the image.}
\label{fig:qubic_schem}
\end{figure}

QUBIC aims to demonstrate the advantages of spectral imaging~\cite{spectral_imaging} in the context of component separation in the milimetre wavelengths. Its beam (or point spread function, PSF) is an interference pattern created by images of the sky through an array of 64 back-to-back feedhorns. It results in a multi-lobed beam, which means that QUBIC observes different directions in the sky at the same time (Fig.~\ref{fig:qubic_schem}). The distance between higher order lobes and the order~0 is inversely proportional to the frequency of the incident light. The absolute positions and amplitudes of the lobes must be calibrated for each of the 248 transition edge sensors (TES), the detectors on the focal plane. With this knowledge of the QUBIC beam, it is then possible to obtain the spectrum of each pixel when reconstructing a map of the sky, in a similar manner to Euclid NISP instrument with its grism.

This is a strong advantage of QUBIC with respect to direct imagers: it does not need many frequency bands and focal planes to attain a high frequency resolution. The foregrounds are not yet fully understood and numerous models could describe their spectral signatures, with varying complexity. In order to perform a component separation with complex models, a higher frequency resolution is necessary.

\section{Map making methods}

The map making step transforms time-ordered-data (TOD) into maps of the sky. These maps can be ``snapshots'' of the sky observed at different frequencies or they can be maps of the distributions of the components that emit light at those frequencies (CMB, dust, synchrotron, etc.). Classical map making links the TOD to the sky with the following equation:

\begin{equation}
	\vec{d} = H_{\nu} . \vec{s_{\nu}} + \vec{n},
	\label{eq:mapmaking}
\end{equation}

\noindent
where $\vec{d}$ is the TOD, $H_{\nu}$ is the instrument model including the bandwidth and the pointing matrix (the position of each TOD on the sky), $\vec{s_{\nu}}$ is the sky at the observed frequency and $\vec{n}$ is the noise. In most cases $H$ is not invertible and it is not trivial to retrieve $\vec{s_{\nu}}$. A Preconditioned Conjugate Gradient (PCG) method can be used to approach a good estimation of $\vec{s}$ by iterating on a sky realisation and comparing the associated TOD with the real TOD. The QUBIC collaboration has developed more advanced map mapking that take advantage of the spectral imaging in order to extract more information from the TOD.

\subsection{Frequency map making}

The first one is the Frequency Map Making~\cite{FMM} (FMM). The idea is to build more than one frequency map from one TOD using spectral imaging. Eq.~\ref{eq:mapmaking} becomes

\begin{equation}
	\vec{d} = \sum_{i=1}^{N_{\rm sub}} H_{\nu_i} .  \vec{s_{\nu_i}} + \vec{n},
	\label{eq:FMM}
\end{equation}

\noindent
where the frequency band corresponding to the TOD is split in $N_{\rm sub}$ sub-bands at frequencies $\nu_i$. $ H_{\nu_i}$ and $s_{\nu_i}$ are respectively the instrument model and the sky at frequency $\nu_i$. Using PCG it is then possible to retrieve several $s_{\nu_i}$ from only one TOD. This allows QUBIC to reach a high frequency resolution that will help with component separation.

\subsection{Component map making}

The frequency maps step can be skipped by doing the component separation at TOD level. This approach is called Component Map Making~\cite{CMM} (CMM) and it fully takes advantage of spectral imaging. In this context, Eq.~\ref{eq:FMM} becomes

\begin{equation}
	\vec{d} = \sum_{i=1}^{N_{\rm sub}} H_{\nu_i} .  A_{\nu_i}\vec{c} + \vec{n},
	\label{eq:CMM}
\end{equation}

\noindent
the frequency maps $\vec{s_{\nu_i}}$ are decomposed into a mixing matrix $A_{\nu_i}$ and component maps $\vec{c}$. It is then possible to fit the mixing matrix along with the component maps, which means we can still obtain the frequency maps but the component separation is already performed during the PCG step.

QUBIC beam has multiple lobes, the order 1 being about 10 degrees away from the order 0. When pointing near the edge of QUBIC patch ($\sim1.5\%$ of the sky), the side lobes will fall in a region not well observed. This results in a degeneracy in the PCG reconstruction of the sky near the edge of the patch, that we mitigate by adding \textit{Planck} data outside the patch.

\subsection{Neural network map making}

The PCG is numerically expensive and does not use perfectly the pointing information. The Neural Network Map Making approach~\cite{NNMM} is a physics guided modular neural network. It inverts each separate component of $H_{\nu}$ (pointing, filters, etc.) with a sub neural network with parameters chosen to fit their theoretical invert. It results in a flexible and interpretable approximation of $H_{\nu}^{-1}$ that does not need a full retraining if one component of the instrument is changed. It means that once trained, the map making step becomes trivial. This method showed better results than FMM and CMM as it properly handles the complex pointing matrix of QUBIC.

\section{Moon maps}

QUBIC has resumed its commisioning and has recently observed the Moon. We want to measure the Moon spectrum on maps without beam deconvolution as a proof of concept of QUBIC spectral imaging. We present here maps of the Moon obtained by scanning at a constant elevation $\textrm{el}=50~\deg$ and an azimuth range $\Delta \textrm{az} = 135 - 85 =  50~\deg$. The observations last for approximately 2~hours, the time for the Moon to move 25 degrees vertically in the sky. With this $25\times50~\deg^2$ range, the Moon crossed all order 0 lobes of all TES beams and most of the order 1. It is used as a first calibration of the lines of sight of all TES as well as a their gains.

In order to obtain the beam from the image of the Moon through the beam, we need to use the right set of coordinates that show the angular distance from the order 0 Moon's image and the direction with respect to the beam horizontal line, corrected for the Moon's movement. This correction is also necessary on a static source, such as the calibration source used for the beam lab measurement on Fig.~\ref{fig:qubic_schem} . 

\begin{figure}
	\includegraphics[width=0.37\textwidth]{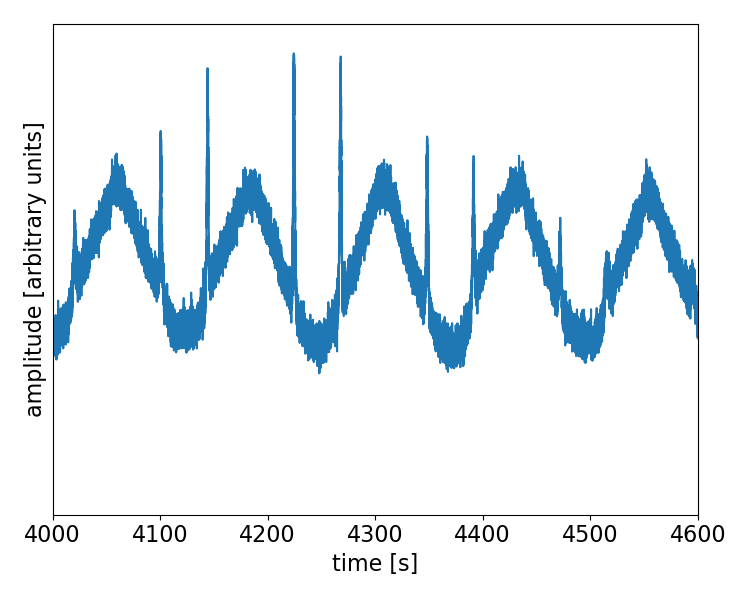}
	\includegraphics[width=0.3\textwidth]{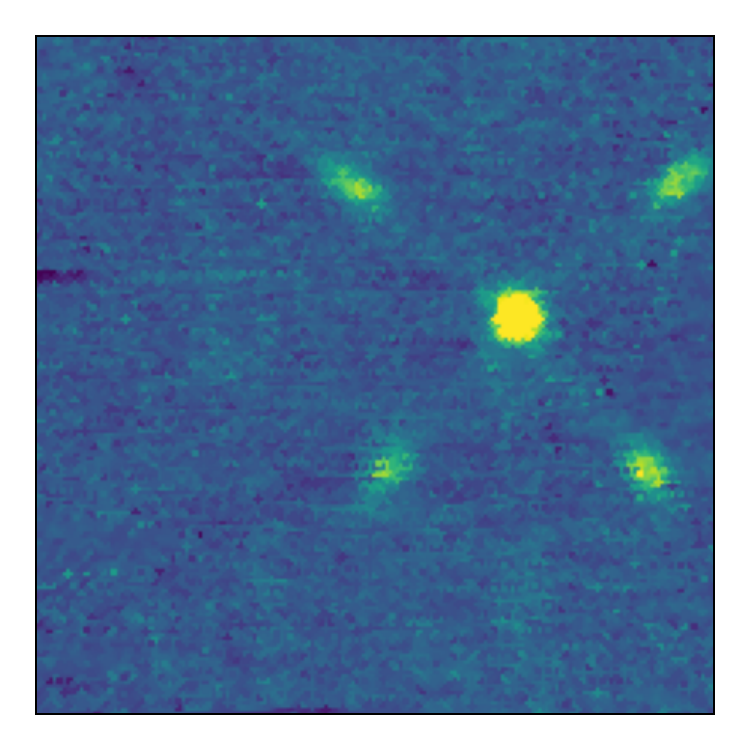}
	\includegraphics[width=0.3\textwidth]{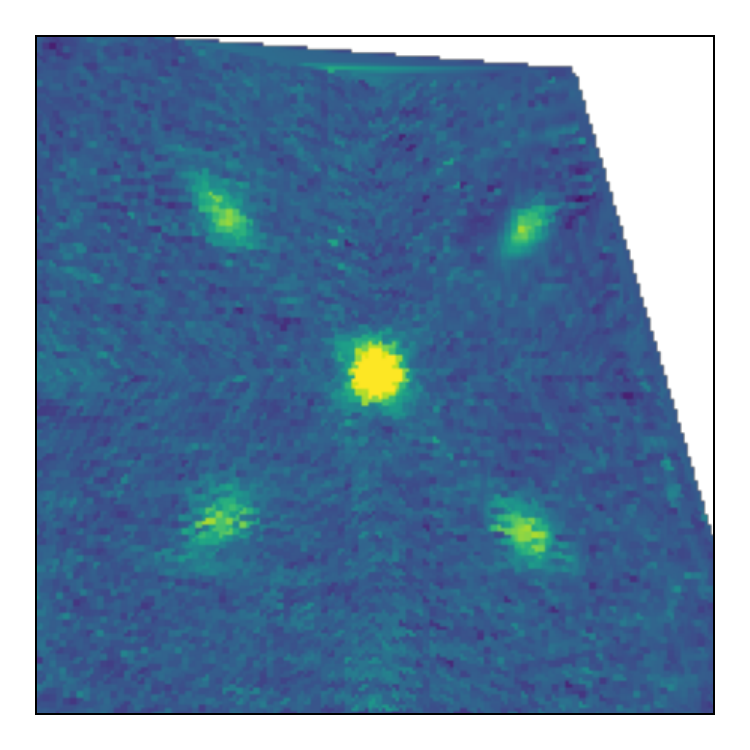}
	\caption[]{\textit{Left:} raw TOD for one TES when the Moon is at the elevation of its order 0 lobe. See the text for more details. \textit{Middle:} $25\times25~\deg^2$ map of the Moon for one TES, only corrected for the Moon's movement. \textit{Right:} $25\times25~\deg^2$ map of the Moon for one TES with the coordinates correction. The shift of the order 0 Moon image is simply due to the coordinates definition.}
	\label{fig:moon}
\end{figure}

Fig.~\ref{fig:moon} shows, for one TES, a small portion of the TOD and the effect of the coordinates correction on the shape of the Moon's image. The large oscillations in the TOD follow the azimuth variation and are due to the magnetic field of the Earth impacting the SQUIDs. This has been confirmed by observations made of the same field but with the instrument window covered so that sky and atmosphere are not contributing to the measurement. The small spikes are the Moon and the rest is instrumental and atmospheric noise.

\section{Prospects}

The next step will be to fit a combination of QUBIC beams for a number of sub-bands, similar to the FMM approach but on a quasi-point source which means the analysis can be done at map-level in this context. The fitting parameters should then depend on the Moon spectrum but also the atmosphere emission and transmission in a given sub-band. At the moment, we plan to use the zone without Moon signal to measure the atmosphere emission and deduce its transmission.

\section*{References}
\bibliography{moriond}


\end{document}